# Deep Neural Network Voice Activity Detector for Downsampled Audio Data: An Experiment Report

1st Mikael Ovaska
*Faculty of Information Technology*
*University of Jyväskylä*
Jyväskylä, Finland
mikael.t.ovaska@jyu.fi

2nd Joni Kultanen
*Faculty of Information Technology*
*University of Jyväskylä*
Jyväskylä, Finland
joni.m.kultanen@jyu.fi

3rd Teemu Autto
*Faculty of Information Technology*
*University of Jyväskylä*
Jyväskylä, Finland
teemu.a.autto@jyu.fi

4th Joonas Uusnäkki
*Faculty of Information Technology*
*University of Jyväskylä*
Jyväskylä, Finland
joonas.t.uusnakki@jyu.fi

5th Antti Kariluoto
*Faculty of Information Technology*
*University of Jyväskylä*
Jyväskylä, Finland
antti.j.e.kariluoto@jyu.fi

6th Joonas Himmanen
*Faculty of Information Technology*
*University of Jyväskylä*
Jyväskylä, Finland
joonas.e.himmanen@jyu.fi

7th Mikko Virtaneva
*Workspace Oy*
Helsinki, Finland
mikko.virtaneva@workspace.fi

8th Pasi Kaitila
*Workspace Oy*
Helsinki, Finland
pasi.kaitila@workspace.fi

9th Pekka Abrahamsson
*Faculty of Information Technology*
*University of Jyväskylä*
Jyväskylä, Finland
pekka.abrahamsson@jyu.fi

*Abstract*—Sociometric badges are an emerging technology for study how teams interact in physical places. Audio data recorded by sociometric badges is often downsampled to not record discussions of the sociometric badges holders. To gain more information about interactions inside teams with sociometric badges a Voice Activity Detector (VAD) is deployed to measure verbal activity of the interaction. Detecting voice activity from downsampled audio data is challenging because down-sampling destroys information from the data. We developed a VAD using deep learning techniques that achieves only moderate accuracy in a low noise meeting setting and in across variable noise levels despite excellent validation performance. Experiences and lessons learned while developing the VAD are discussed.

## I. INTRODUCTION

Studying how teams work and interact is an interesting and emerging field of study. In a push towards real-time data collection and analysis of human behavioral data, many systems based on wearable devices with different sensor modalities have been developed [1]. Sociometric badges are emerging wearable technology that is being developed to study how teams interact in places. For example, in the Rhythm system developed by Lederman et al., which is developed to measure interaction inside a team, sociometric badges are utilized [2]. The Rhythm system can collect interaction data in co-located formal meetings, co-located informal meetings, distributed remote meetings, and hybrid meetings. To collect voice data in co-located and hybrid meetings sociometric Rhythm badges are used [2-3]. Rhythm badges collect two types of data: vocal activity and proximity to other badges. By combining these three measures it is possible to determine who interacted with each other, what locations interactions happened in, and what were the dynamics of the interactions. Interaction dynamics are attained by analyzing total times of speech, average lengths of speech, turn-taking [4], or overlap of speeches inside a discussion. The system relies heavily on Voice Activation Detector (VAD) to create the analysis, making VAD a critical part of the system. The accuracy of it influences almost all the analysis created from the system's data. Sociometric badges like Rhythm badges often downsample the data sent from badges to provide privacy for the discussion of the users. Since the data used for VAD is downsampled many of the common features for VAD are not available (analysis of features for VAD reviewed at [5]) and therefore all the advancements of VAD technology are not available in this context.

Since VAD is a critical piece in a puzzle to measure interaction with sociometric badges, it deserves scientific attention. Current VAD methods for sociometric badges are very simplistic and rough due to the inability to use the most advanced VAD techniques. We took a challenge to develop

VAD for sociometric badges further to empower the sociometric badge ecosystem to study team interaction more effectively. Deep learning methods have been shown to work well in VAD with input that is not downsampled [6], so we decided to try to implement a VAD with deep learning. We trained a binary classification model to detect speech from audio volume data collected at 20Hz frequency. Audio volume data was collected with Rhythm badges developed by MIT [2-3].

## II. BACKGROUND

A voice activity detector (VAD) is an algorithm that attempts to separate the audio stream into intervals where voice activity is present and where it is absent [5]. Commonly VAD methods used with sociometric badges use empirically found constant threshold to classify voice activity [2-3][7-9]. Threshold values in these methods depend on the usage environment and it is hard to choose the threshold objectively since there aren't many tools to evaluate the correctness of the threshold value. The lower bound of the threshold values can be found by inspecting values in a period that is known to have no voice activity, but then it is up to the user to choose the threshold. J. Shen et al. [9] used VAD that relies on the idea that the speech of one person in a meeting will be recorded by other persons' badges microphone too. Then if recordings of meeting participants correlate a lot, it is likely that there is voice activity in the recording at that moment and the recording with the highest volume is the most to likely contain the genuine voice activity. In that VAD there is still one choice left that is hard to objectively choose for the user, the level of correlation that is as a voice activity by the algorithm. That problem could be solved by studying and validating the correct level of correlation, but the algorithm would still only be able to detect one speaker's voice activity at a time.

## III. EXPERIMENTAL SETTING

The sociometric badge system described and used in [2-3][8] was used. The system has been called by two different names, "Rhythm" and "Openbadges". We will reference it as a "Rhythm" when talking about the system in general and with the "Openbadge" name when referencing a specific code repository that goes by that name. Systems specifications and used code is listed in Table 1. Rhythm badges don't implement VAD in the hardware, so it is done in the backend of the system. Badges sample microphone signal at 700 Hz and creates an average amplitude reading every 50 ms which is sent to the backend [2]. No audio data is collected, so no conversations are sent to the backend providing privacy for the users. The project followed the typical workflow of developing a machine learning model [10].

### A. Model definition

The first requirements for the model were defined. The scope of the model was defined to detect speech from data produced by Rhythm badges in a static offline office meeting. Relatively real-time visualization (less than 10-second delay) of the VAD is desirable, so VAD must work with a relatively small amount of information. The VAD should not have any parameters that need to be configured to fit the usage environment.

### B. Data collection

Since open-labeled Openbadge or Rhythm datasets were not available, we planned and carried on a meeting to gather data. A rhythm standalone server was set up in the office to collect the data produced by badges. The office room where the meeting was held was quite large and open room. The physical setup of the meeting was as shown in "Fig 1.". Minimum distances between participants were controlled to be close to equal, every participant was at least 1 meter away from other participants. The meeting involved six participants, 5 male and 1 female. The first half of the meeting was held as a "normal" civilized meeting between participants. Each participant spoke on their own turns after each other. The second half included more dynamic meeting settings that could happen while workshopping or brainstorming, and rare situations and edge cases that could occur in a meeting to the dataset. These scenarios included multiple 1-on-1 discussions happening simultaneously, loud environmental noises (door opening or closing, sound of hitting a table, music video playing in television), speaking while standing, speaking while walking in the room, and interrupting speaker on purpose. Both English and Finnish language were used in the meeting. The meeting was filmed with a video camera to collect the ground truth. Participants were introduced to raise a colormaker every time they speak to make labelling easier. The meeting was kicked off by clapping for 15 seconds.

Test data was gathered at a separate meeting after the model selection was done. By gathering the testing dataset after developing the models, data leakage from test data to the model isn't possible, and therefore a better understanding of the performance of the model can be attained. The meeting was designed to be slightly different than the meeting where training data was gathered to get some idea of how presentative the training dataset is to the general population, and how well the selected model could generalize to a slightly different setting. The second meeting had 8 participants, seven male, and one female. Five of the male participants took part in the training data gathering meeting too. The language of the meeting was Finnish. All participants were advised to start a recording also on their mobile phones to two ensure that good quality audio of the discussions would be available. Data of three participants was selected to be annotated for the model's performance testing. These three participants will be called later $x$, $y$, and $z$. Participant $x$ had participated in the training data gathering meeting too. Male participant $y$ and female participant $z$ participated for the first time. The physical setup of the meeting is shown in "Fig 2.". The meeting was split into 5 different phases; 1) Civilized meeting, one person speaking at the time 2) 1-on-1, four simultaneous discussions at the same time 3) 1-on-1 with the television on to increase the background noise 4-5) 1-on-1 seats switched to change the audio environment of the discussion's television on and off. As environment noise football talk show in the English language was played on the television. The meeting was kicked off by jointly clapping for 15 seconds to create an artificial spike to the audio data. Between phases, there were some discussions, which were also annotated.

TABLE I. SOCIOMETRIC BADGE SYSTEM'S SPECIFICATIONS. SPECIFIC COMMIT SHA-1 IS AFTER THE LINK TO REPOSITORY

| Sociometric badges | MIT HumanDynamics Openbadge https://github.com/HumanDynamics/openbadge/ #a72b1ad95fbb642cdc694f44d654276fd82046c2 |
|---|---|
| Badge's casing | 3d printed case |
| Microphone sampling | 700 Hertz |
| Badge audio output | Average amplitude reading every 50 milliseconds (20 Hertz) |
| Badge hub | Openbadge hub https://github.com/HumanDynamics/openbadge-hub-py #d80eaa3293c61b78e19379d0db258281b1a78e85 |

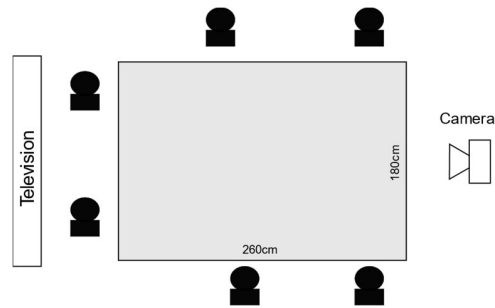

Fig 1. Training data gathering meeting

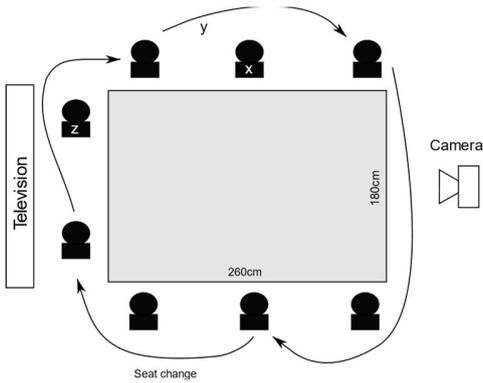

Fig 2. Test meeting setup

*C. Data annotation*

After collecting the dataset, it was labelled using the video record as a ground truth. When the speech of a participant was too hard to hear from the video, audio recorded by the participant was added to the video. Intervals, where voice activity was noticed, were marked by writing down the beginning and end of the interval. Annotation was as accurate as possible to include even short intervals of activity and then rounded to every 50ms. Different persons annotated training and test datasets.

*D. Feature engineering*

To prepare data to be used to train the models, feature engineering was done. First, the raw audio volume data was converted to pivot table in pandas data frame format [11] with sample2data function provided in Openbadge repository by MIT. Pivot table for audio data follows the same logic as pivot table for speech labels, index of the table is time, badges are columns and cells contain the audio volumes collected by badges.

The next step was to create samples out of devices that produced a continuous stream of data. Samples were created by taking samples with a rolling window over time. Three second time window was chosen.

*1) One channel*

To build the simplest possible solution at first, only the primary badges, whose voice activity is to be detected, own data was used as an input for the model. For the rest of the paper, this version will be referred to as 'one channel solution' since the model has only one channel of the input source.

*2) Multiple channels*

In the next iteration, we wanted to add more data sources to the model. We theorized that, if the mean of the badge's audio volumes changes a lot when the primary badge is removed, it is likely that the audio source is close to the primary badge. When the primary badge is worn in the neck the audio source is very likely to be the mouth of the wearer. Therefore, that difference between means could be a good input feature to use in VAD. This was also supported by analyzing boxplots of that feature on two labels. When 3 second rolling mean of mean differences is applied, a clear difference in distributions is seen between two labels. Distributions of 3 second rolling mean of standard deviation difference in the same logic on the two labels are also different, so it was included as a feature. Audio data was included in the model as a feature as well because by looking at the boxplot a slight difference is seen in the distributions of labels. 3 second rolling mean of variance difference is also included as a feature for model selection, even though there is no clear difference in the distributions between labels. Boxplots of features are in "Fig 3.".

*3) Normalisation*

The requirement to make detections relatively quickly adds constraints for normalization. Since the result of the VAD needs to be ready for visualization in a short amount of time samples cannot be normalized for a long period, for example for the whole period of the meeting. Therefore, normalization is best to be done inside the input samples. Then normalization is independent for every prediction the model makes. Models without any normalization in preprocessing and L2 normalization were included for the model selection process.

*E. Model selection and performance estimation*

Stratified 5-fold cross-validation with an independent test set was used as a method to select a model and estimate its performance. The test set was gathered from the meeting held after the model selection was done. The model with the best average balanced accuracy on cross-validation was selected as the best model. The best model was then trained on the whole training set and its performance in the test set was evaluated by looking at balanced accuracy and F1 score [12-13] at different phases in the test set.

*F. Model training*

All models were built and trained using Keras [14]. Class weights were calculated using the scikit-learn python library [15] and binary cross-entropy was used as a loss for all models. Each fold was run for 15 epochs with a batch size of 4000. The final training of the selected model was run until binary accuracy converged.

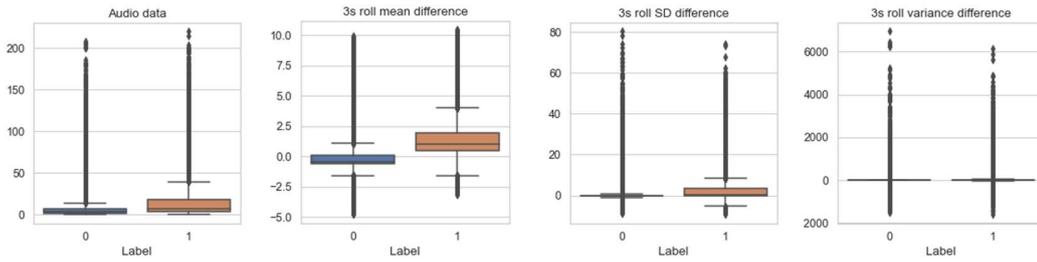

Fig 3. Distributions of selected features. "0" = No Voice activity, "1" = Voice Activity

## IV. Key lessons learned

In what follows, several key lessons learned are identified using Primary Empirical Observation (PEO) as the concept [16].

### A. PEO 1: Mistakes at preprocessing might not be detectable in the validation of the model.

At the beginning of the project, the mistake of not aligning the labels table and the audio data table columns correctly was made. The models were trained with a dataset, where labels of the data were not correctly set to corresponding audio data. Ie. in the audio table column order was A, B, C, D when in label table columns order was C, D, A, B. As consequence, the model trained to detect when person C spoke using data from the badge of person A. And another coding mistake that happened in the project was not aligning the audio table and labels table to begin and end at the same time. The badges were turned on before the video recording started, but in the code audio stream and video stream were assumed to begin at the same time, which resulted in audio and targets to be not synced, targets being about 20 minutes behind the volume.

In neither of the cases, mistakes were found because of models' bad validation accuracy. Models fitted well in the training and validation sets even if the input data and targets were not correctly aligned. Because the targets of the model were wrong, the model would not have generalized at all, so without finding these mistakes we would have been disappointed in the model's predictive ability in the test set.

### B. PEO 2: The generalization ability of the model can be evaluated visually even with data without ground truth.

The inability of the 'one channel solution' to detect the speaker correctly was tested by applying it to the dataset gathered in the office-wide experiment. The dataset was not labelled since no ground truth for speech was gathered in the experiment. Even though the dataset is not labelled, it could be used to evaluate VAD performance roughly. 'One channel solution' predicted that voice activity occurred often at the same time in all badges. When looking at patterns at those times, we noticed that the same pattern was present at all the badges, but with different magnitudes, indicating that the model cannot distinguish is the voice activity from the primary badge or not.

### C. PEO 3: Creating a simultaneous artificial spike in the training data to sync data from different sources helps in the annotating process.

We began the data gathering experiments by clapping together to create a simultaneous artificial spike in the volume data in all badges data to sync the time between different sources of data. This clapping is then seen in the video we recorded too, even though we had a clock pointing to the camera too in the later meeting. Clapping had in our experiment the same function as sync slate in TV and movie production. By using clapping in front of the badge to synchronize different sources, we made sure that the spike is large enough to be noticed. This practice helped us in the annotation process and in situations where time deviation was suspected. For example, our sociometric badge system's time records deviated from the true time for more than 1 minute in the test set, and we were able to correct the data easily thanks to having an artificial spike.

### D. PEO 4: Definition of Voice Activity changes the annotation of audio data and therefore performance evaluation noticeably.

The reason VAD is deployed in sociometric badge context is to gain information about verbal interaction. The definition of voice activity affects how the VAD system is designed and how its performance is evaluated. Are filler sounds voice activity? Should a half-second pause in the speech be labeled as voice activity or not? What should be the graduality of the predictions, should predictions be done every second or 50 milliseconds? VAD systems that have different definitions to these questions for example will attempt to detect different things, making the straight comparisons of two systems' performances difficult.

### E. PEO 5: Annotating audio data accurately can be challenging, which affects model evaluation.

Accuracy of the annotation has most certainly an effect on the model and its test performance. At the same time, the manual annotation process when done preciously will be a tedious and long process prone to errors, because the gap between speech turns can be as short as 250ms [17]. Annotating the voice activity of 6 participants in a 1-hour meeting even at half-second accuracy can be a task that takes one week of full workdays to complete. If errors are made when annotating the test set, the model could be predicting correctly, but it would be evaluated as a mistake. With short words such a mistake can be easily done, missing a half-second long activity by half second is a 0% overlap. Even when not using supervised machine learning annotation is needed to evaluate the performance of the VAD.

*F. Practical implications*

*1) It is the model's designers responsibility to make sure that the model is learning a correct thing*

Deep learning networks have a great ability to succeed in the problems their trainer gives it, but trainers must make sure that the problem makes sense, and the problem is correctly presented to the network. In this project, the code defects were often found when the progress of the work needed to be presented to others. This implies that common practices to enhance the code quality and catch defects from software engineering such as code reviews [18-19] should be used in machine learning projects. Also, the annotation process needs to be consistent in both training and test datasets.

*2) Not only quantitative methods can be used to evaluate machine learning model's performance*

The performance of the model can be evaluated roughly, and code defects can be found, by visualization of the predictions, but metadata needs to be available to make use of visualization well. For example, the deviation in time between systems could be very hard to spot without effort to create artificial spikes in the audio data.

*3) Possibilities for errors should be identified and measures to prevent them should be taken at the beginning of the project*

Famously Murphy's law states that "whatever can go wrong, will go wrong.". When developing a machine learning system, making mistakes and errors is easy in all stages of the project. These errors could for example lead to overestimating the predictive power of the model, which could result in costly disappointment later when the model is being implemented. Therefore, measures to remove possibilities of errors should be taken whenever feasible. Measures could be anything from using code reviews to gathering test data only after model selection. To have oversight to see possible errors, a team needs to have knowledge and experience about software engineering, machine learning, and the domain. To spread the knowledge about the errors, researchers and practitioners must document and publish them, for example in a form of an experiment report.

TABLE II. MODEL TYPES

|  |  | CNN |
|---|---|---|
|  |  | Input layer (60,3) |
|  |  | Conv1D, filter=254, kernel=3, padding=same |
| 4x |  | Batch normalization |
|  |  | ReLu activation |
|  |  | Global average pooling |
|  |  | Sigmoid activation |
|  |  |  |
|  |  | LSTM+LSTM |
|  |  | Input layer (60,3) |
| 2x |  | LSTM, units=100 |
|  |  | Sigmoid activation |
|  |  |  |
|  |  | CNN+LSTM / CNN+LSTM+LSTM |
|  |  | Input layer (60,3) |
|  |  | Conv1D, filter=64, kernel=4, padding=same |
|  |  | ReLu activation |
|  |  | Max pooling, pool size=2 |
| CNN+LSTM+LSTM 2x |  | LSTM, units=100 |
|  |  | Sigmoid activation |

TABLE III. 5-FOLD CROSS VALIDATION SCORES FOR MODEL SELECTION

| Model type | Features | Normalization | Cv val score | Cv train score |
|---|---|---|---|---|
| CNN | A | No | 0.948 | 0.962 |
|  |  | L2 | 0.954 | 0.960 |
|  | B | No | 0.947 | 0.969 |
|  |  | L2 | 0.945 | 0.969 |
| CNN+LSTM | A | No | 0.962 | 0.966 |
|  |  | L2 | 0.932 | 0.938 |
|  | B | No | 0.964 | 0.967 |
|  |  | L2 | 0.931 | 0.940 |
| CNN+LSTM+LSTM | A | No | 0.972 | 0.976 |
|  |  | L2 | 0.941 | 0.949 |
|  | B | No | **0.973** | 0.978 |
|  |  | L2 | 0.938 | 0.945 |
| LSTM+LSTM | A | No | 0.967 | 0.973 |
|  |  | L2 | 0.931 | 0.939 |
|  | B | No | 0.964 | 0.969 |
|  |  | L2 | 0.928 | 0.935 |

## V. EXPERIMENT RESULTS

Four different types of models were included in the model selection (TABLE II). Two different sets of features were included in the model selection; "A" has all the features presented in "Fig 3." and "B" has variance difference removed from the feature set. Models without any normalization and with L2 input sample normalization were included. In the results TABLE III "Cv val score" refers to mean balanced accuracy in validation folds and "Cv train score" to mean binary accuracy achieved in the training folds. CNN+LSTM+LSTM with "B" feature set and without normalization achieved the highest mean validation score. The scores between "A" and "B" feature sets are very close to each other, but the "B" set would be preferred over "A" because the number of features should be minimized if possible. Overall models without normalization outperformed models with L2 input sample normalization.

TABLE IV. TEST SCORES

| Scenario | Subject | Balanced accuracy | F1-score |
|---|---|---|---|
| Normal Meeting | x | 0.827 | 0.572 |
|  | y | 0.675 | 0.500 |
|  | z | 0.918 | 0.779 |
|  | Overall | 0.800 | 0.611 |
| 1 on 1 | x | 0.680 | 0.551 |
|  | y | 0.618 | 0.398 |
|  | z | 0.575 | 0.401 |
|  | Overall | 0.610 | 0.441 |
| 1 on 1 seats switched | x | 0.689 | 0.557 |
|  | y | 0.637 | 0.482 |
|  | z | 0.548 | 0.308 |
|  | Overall | 0.615 | 0.445 |
| 1 on 1 TV on | x | 0.765 | 0.703 |
|  | y | 0.680 | 0.552 |
|  | z | 0.635 | 0.635 |
|  | Overall | 0.688 | 0.580 |
| 1 on 1 TV on seats switched | x | 0.720 | 0.629 |
|  | y | 0.635 | 0.440 |
|  | z | 0.503 | 0.328 |
|  | Overall | 0.616 | 0.458 |
| Whole meeting | x | 0.728 | 0.568 |
|  | y | 0.644 | 0.644 |
|  | z | 0.638 | 0.638 |
|  | Overall | 0.666 | 0.488 |

TABLE V. TEST SCORES

| True labels | Predicted labels | | | | | |
|---|---|---|---|---|---|---|
| | **Normal Meeting** | | | | | |
| | x | | y | | z | |
| | 0 | 1 | 0 | 1 | 0 | 1 |
| 0 | 16909 | 2236 | 19235 | 162 | 19371 | 574 |
| 1 | 553 | 1862 | 1387 | 776 | 218 | 1397 |
| | **1 on 1** | | | | | |
| | x | | y | | z | |
| | 0 | 1 | 0 | 1 | 0 | 1 |
| 0 | 4750 | 455 | 2911 | 46 | 3738 | 658 |
| 1 | 1411 | 1144 | 3599 | 1204 | 2354 | 1010 |
| | **1 on 1 seats switched** | | | | | |
| | x | | y | | z | |
| | 0 | 1 | 0 | 1 | 0 | 1 |
| 0 | 3138 | 215 | 1482 | 78 | 1921 | 216 |
| 1 | 842 | 665 | 2228 | 1072 | 2187 | 536 |
| | **1 on 1 TV on** | | | | | |
| | x | | y | | z | |
| | 0 | 1 | 0 | 1 | 0 | 1 |
| 0 | 1961 | 98 | 1470 | 44 | 1725 | 167 |
| 1 | 635 | 866 | 1250 | 796 | 1071 | 597 |
| | **1 on 1 TV on and seats switched** | | | | | |
| | x | | y | | z | |
| | 0 | 1 | 0 | 1 | 0 | 1 |
| 0 | 2614 | 139 | 2229 | 38 | 1709 | 510 |
| 1 | 1022 | 985 | 1780 | 713 | 1943 | 598 |
| | **Whole meeting** | | | | | |
| | x | | y | | z | |
| | 0 | 1 | 0 | 1 | 0 | 1 |
| 0 | 43765 | 6011 | 46015 | 531 | 47337 | 3496 |
| 1 | 5624 | 7660 | 11572 | 4942 | 8018 | 4209 |

The test set model achieves 0.8 balanced accuracy and 0.611 F1-score in the normal meeting setting, 0.61-0.688 balanced accuracy, and 0.411-0.58 F1-score in the 1-on-1 discussions and for the whole meeting balanced accuracy 0.666 and F1-score 0.488. Interestingly the model performed better when TV was on than off. Scores are seen in TABLE IV. The confusion matrices of predictions are shown in TABLE V. Predictions were done for every 50ms. When observing the predictions closer with confusion matrices, we notice that in the normal meeting setting the model can predict when voice activity is absent well, but it has difficulties predicting when voice activity is present. In 1-on-1 setting models precision is good, but recall is bad.

## VI. IMPLICATIONS

The model developed could generalize well to the test set even though no indication of overfitting was noticed in the 5-fold cross-validation. That indicates that the training set gathered doesn't present the general population well. It is also possible that annotations between training and test data sets differ enough to make performance estimation in the test set be underestimated. Also, balanced accuracy seems to be a bad indicator of the model's performance. For example, the model achieved balanced accuracy of 0.827 for participant x in a normal meeting setting, but when looking at the confusion matrix model predicted that voice activity is present incorrectly more often than correctly. Just looking at balanced accuracy one would think the model is 82.7% accurate, which would overestimate its skill. We didn't explore the hyperparameter and feature spaces completely, so there are most likely many ways to improve the VAD.

## VII. CONCLUSIONS

Sociometric badges are an emerging technology to study how teams interact in physical places. In this paper, we described our experiences and results of developing more advanced VAD with deep learning for sociometric badges. Our VAD achieved an excellent performance in validation data, but in test data, its performance was much lower which is suspected to be because of the training set being badly presentative to the general population. Further research on features and models VAD for downsampled audio data is needed to develop sociometric badges as a tool to study team interaction.